\documentclass[lettersize,journal]{IEEEtran}
\IEEEoverridecommandlockouts
\usepackage{graphicx}    
\newtheorem{theorem}{\bf{Theorem}}[section]

\newtheorem{remark}{\bf{Remark}}
\newtheorem{assumption}{\bf{Assumption}}
\usepackage{amsmath,amssymb,amsfonts}
\usepackage{algorithmic}
\usepackage{algorithm}

\usepackage{array}
\usepackage{textcomp}
\usepackage{stfloats}
\usepackage{url}
\usepackage{verbatim}
\usepackage{graphicx}
\usepackage{cite}
\def\BibTeX{{\rm B\kern-.05em{\sc i\kern-.025em b}\kern-.08em
		T\kern-.1667em\lower.7ex\hbox{E}\kern-.125emX}}
\usepackage{balance}
\usepackage{booktabs}
\usepackage{subfigure}
\usepackage{epstopdf}
\usepackage{epstopdf}
\usepackage{soul}
\usepackage{epstopdf,cite,textcomp,mathrsfs,color,bm}
\usepackage[x11names]{xcolor}       
\usepackage{caption}
\usepackage[font=footnotesize, labelsep=period]{caption}

\makeatletter
\def\endthebibliography{%
	\def\@noitemerr{\@latex@warning{Empty `thebibliography' environment}}%
	\endlist
}
\makeatother
\begin{document}

\include{header}

\title{Integrated Sensing, Communication, and Computation Over-the-Air
in OFDM Systems}
		\author{
        Biao Dong,
	Bin Cao, \IEEEmembership{Member, IEEE},
        and Qinyu Zhang, \IEEEmembership{Senior Member, IEEE}
		\thanks{}
		\IEEEcompsocitemizethanks{
        \IEEEcompsocthanksitem This work was supported in part by the Shenzhen Science and Technology Program under Grant KJZD20240903095402004.
			\IEEEcompsocthanksitem Biao Dong, Bin Cao, and Qinyu Zhang are with the School of Electronic and Information Engineering, Harbin Institute of Technology (Shenzhen), Shenzhen 518055, China (e-mail: 23b952012@stu.hit.edu.cn; caobin@hit.edu.cn; zqy@hit.edu.cn).
	}}


\maketitle

\begin{abstract}
This work is concerned with integrated sensing, communication, and
computation (ISCC) in uplink orthogonal frequency division multiplexing (OFDM) systems, wherein multiple devices perform target sensing and over-the-air computation (AirComp) simultaneously. We aim to minimize the computational mean squared error (MSE) by jointly optimizing the transmitting vector and the aggregation vector. To tackle the non-convexity of the formualted problem, we develop a two-phase iterative algorithm. Simulations demonstrate that the proposed algorithm can achieve a better sensing-computation trade-off.

\end{abstract}

\begin{IEEEkeywords}
Integrated sensing and communication, over-the-air computation, OFDM, integrated sensing, communication, and computation.
\end{IEEEkeywords}

\section{Introduction}


With the proliferation of emerging applications such as autonomous vehicles, remote healthcare, and industrial automation, the integration of sensing, communication, and computation (ISCC) to meet strict latency and reliability requirements is extremely urgent \cite{wen2024survey}. Specifically, this integration involves three key processes: acquiring information (sensing), sharing information (communication), and processing information (computation). These processes are inherently coupled and compete for shared network resources. In this regard, their joint optimization directly determines the system-level performance.

Most existing studies concentrate on partial integration, such as communication-computation \cite{chen2023over} or sensing-communication \cite{liu2021cramer}, which faces two limitations. Firstly, without jointly optimizing all three processes in a unified framework, network resources cannot be utilized efficiently. Secondly, the performance objectives of individual modules are often not aligned with the system-level goals. This misalignment results in suboptimal performance at the system level. Specifically, \cite{chen2023over} proposes a robust AirComp scheme to support multi-device aggregation. However, sensing is completely ignored and multi-device sensing interference is not modeled. In this regard, power allocation does not account for sensing-related Cramér-Rao lower bound (CRLB) constraints. \cite{liu2021cramer} focuses on maximizing sensing performance subject to communication threshold constraints, but neglects the computation. The most relevant prior work to this paper is \cite{li2023integrated}, which performs joint optimization within a unified framework but is limited to a single-carrier system. However, OFDM has been adopted in many practical communication systems. A recent study also demonstrated that OFDM-based ISAC could achieve the optimal waveform design in terms of the lowest ranging sidelobe \cite{liu2024ofdm}. Hence, it is natural to introduce OFDM to the above unified framework.

\begin{figure}[htbp]    
	\centering
	{\includegraphics[width=0.35\textwidth]{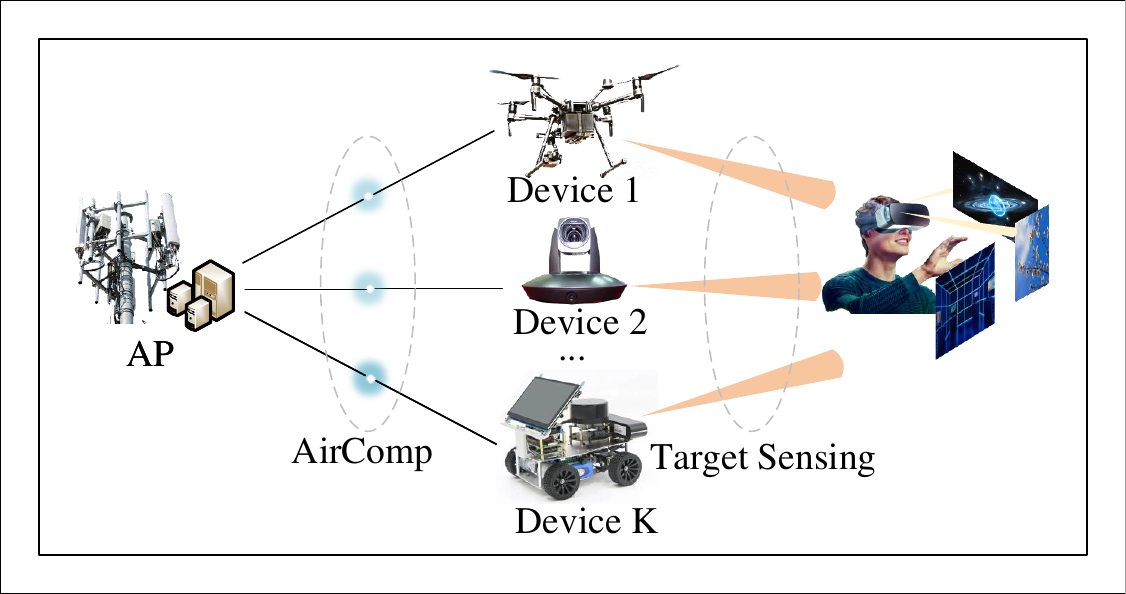}}
	\caption{The considered ISCC system comprises one common target and one AP. Multiple radar devices simultaneously transmit probing signals to detect the target and data symbols to the AP for AirComp. Each radar and AP are equipped with a single antenna.}
    \label{intro_fig}
\end{figure}

In this letter, we extend ISCC to OFDM systems operating over frequency-selective fading channels. We aim to minimize the computational mean squared error (MSE) on all subcarriers via joint optimization of transceiver vectors, subject to CRLB and power constraints. To the best of our knowledge, this is the first work to explore the sensing-computation trade-off under OFDM framework, which presents a research gap within this field. To tackle the formulated non-convex problem, we propose a novel two-phase iterative optimization algorithm, consisting of a successive convex approximation (SCA)-based alternating optimization (AO) phase and an alternating direction method of multipliers (ADMM)-based refinement phase. These two phases respectively focus on decoupling and relaxation of the problem, as well as exploring a broader feasible region. Numerical results show that the proposed method achieves a better sensing-computation trade-off compared to the baselines.

{\textit Notations}: We use lowercase letters to denote time-domain signals (e.g., ${x}_{k,s}$, ${y}_{s}$) and uppercase letters to denote frequency-domain signals (e.g., ${\mathrm C}_{k,n}$, $\mathrm {Y}_{n}$). The vector or matrix is denoted by boldface, $(\cdot)^\top$ denotes the transpose, $(\cdot)^\mathrm{H}$ denotes the conjugate transpose, and $\circ$ denotes the Hadamard product. $\mathcal{CN}(a,\sigma^2)$ denotes the circularly symmetric complex Gaussian (CSCG) distribution with mean $a$ and variance $\sigma^2$.






\section{System Model and Problem Formulation}
\subsection{System Model}
\subsubsection{AirComp Model}
We denote the index of symbol durations by $m \in \mathcal{M}=\{1,\cdots,M\}$ and consider the $m$-th OFDM signal for AirComp 
\begin{equation}
   \mathbf{C}_{k}=\big[\mathrm{C}_{k,0}, \dots, \mathrm{C}_{k,N-1}\big]^\top \in \mathbb{C}^{N\times 1},
\end{equation}
which denotes the 
normalized measured data vector from device $k \in \mathcal{K}=\{0,\cdots,K-1\}$  on $N$ subcarriers\footnote{To simplify the notation, we ignore the subscript $m$ of each subcarrier element $\mathbf{C}^m_{k}$ in $\mathbf{C}_{k}$.}. Let $N$ denote the number of subcarriers and $\mathcal{N}=\{0,\cdots,N-1\}$. Here, $\mathrm{C}_{k}$ is assumed to be independent and identically distributed (i.i.d.) with unit variance and zero mean, i.e., $\mathbb{E}[\mathrm{C}_{k_1,n}\mathrm{C}^*_{k_1,n}] = 1, \mathbb{E}[\mathrm{C}_{k_1,n}\mathrm{C}^*_{k_2,n}] = 0,\forall n\in {\mathcal{N}}, k_1, k_2\in \mathcal{K},k_1 \neq k_2$ \footnote{For the multi-view sensing \cite{wen2024survey}, where multiple devices detect the target from different views, each view can be independent, thus aggregating them allows us to estimate the overall expectation or central tendency in statistics.}. AirComp aims to calculate the average value of all devices on a certain subcarrier as  
\begin{equation}  {\mathrm{F}}_n=\frac1K\sum_{k=1}^{K}\mathrm{C}_{k,n},\quad\forall n\in\mathcal{N}.
\end{equation}
In addition, $\mathbf{X}_{k}=\big[\mathrm{X}_{k,0}, \dots, \mathrm{X}_{k,N-1}\big]^\top\in \mathbb{C}^{N\times 1}$ is the transmitted signal given by $\mathbf{X}_{k}=\mathbf{B}_{k}\circ \mathbf{C}_{k}$ \footnote{This system considers analog encoding, i.e., the transmitted signal at each subcarrier is viewed as a continuous signal instead of a discrete one (discrete-time analog transmission) \cite{shao2022semantic}.}, where $\mathbf{B}_{k}=\big[\mathrm{B}_{k,0}, \dots, \mathrm{B}_{k,N-1}\big]^\top\in \mathbb{C}^{N\times 1}$ is the frequency domain transmitting vector. Each device satisfies the power constraint as follows
\begin{equation}
     \sum_{n=0}^{N-1}\mathbb{E}[|\mathrm{X}_{k,n}|^2]=\|\mathbf{B}_k\|^2\leq P_\mathrm{t},\forall k \in \mathcal{K}.
\end{equation}
Accordingly, the time domain OFDM signal at sample $s$ generated by device $k$ after the inverse discrete Fourier transform (IDFT) is
\begin{equation}\label{eq:Transmitting_OFDM_Signal}
    x_{k,s}=\frac{1}{\sqrt{N}}\sum_{n=0}^{N-1}\mathrm{X}_{k,n}e^{j\frac{2\pi}{N} sn}, \forall k\in\mathcal{K},s\in\mathcal{N}. 
\end{equation} Next, we add cyclic prefix (CP) with length $N_c$ as
\begin{equation}\label{ofdm}
\boldsymbol{x}_{k}=\big[{x}_{k,N-N_c}, \dots, {x}_{k,N-1}\big]^\mathrm{T}\in \mathbb{C}^{\left(N+N_c\right)\times 1}.
\end{equation} 

Assume that the wireless channel from device \(k\) to the AP has a memory of length \(L_{\mathrm{dA}}\), characterized by taps \(\{h_{k,1}, \dots, h_{k,L_{\mathrm{dA}}}\}\)\footnote{Here, the channel state information (CSI) can be obtained at the AP and the sensors via channel estimation by exploiting the channel reciprocity. }. Accordingly, the received signal can be expressed as
\begin{equation}
    {y}_s=\sum_{k=0}^{K-1}\sum_{l=0}^{L_\mathrm{dA}-1}{h}_{k,l}x_{k,s-\tau_{k,l}}+{ \omega}_s,\forall s\in\mathcal{N},  
\end{equation}
where $\tau_{k,l}$ denotes the delayed arrival of the $l$-th path and ${\omega}_s \sim  \mathcal{CN}(0,\sigma_\omega^2)$ denotes the noise at the AP. Performing discrete Fourier transform (DFT) on the received signal, we have
\begin{equation}
    \mathrm Y_n=\sum_{k=0}^{K-1}\mathrm H_{k,n}\mathrm{B}_{k,n}\mathrm{C}_{k,n}+ \Omega_n, \forall n\in\mathcal{N},
\end{equation}
where ${\Omega_{n} = \frac{1}{\sqrt{N}}\sum_{n=0}^{N-1}{\omega}_se^{-j\frac{2\pi}{N} ns}}$ denotes the noise at each subcarrier with distribution $\mathcal{CN}(0,\sigma_\omega^2)$ and ${\mathrm H_{k,n}= \frac{1}{\sqrt{L}}\sum_{l=0}^{L_{\mathrm{dA}}-1}{h}_{k,l}e^{-j\frac{2\pi}{N} ln}}$. Next, the AP performs the post-processing using the aggregation vector $\mathbf{W}=\big[\mathrm{W}_{1}, \dots, \mathrm{W}_{N}\big]^\top\in \mathbb{C}^{N\times 1}$ as
\begin{equation}  \label{8}
\hat{\mathrm{F}}_n=\frac1K\mathrm{W}_n{\mathrm Y}_n,\quad\forall n\in\mathcal{N}.
\end{equation}
where $\hat{\mathrm{F}}_n$ is the average value computed from the signals received by the $n$-th subcarrier aggregation coefficient $\mathrm{W}_n$ across $K$ devices. 

Finally, the MSE between the ground truth $\mathrm{F}_n$ and the estimated value $\hat{\mathrm{F}}_n$ is computed to quantify the computation performance, which is defined as
\begin{footnotesize}
\begin{equation} \label{obj}
    \begin{aligned}
        \mathrm{MSE} & = \mathbb{E}\left[\frac{1}{N}\sum_{n=0}^{N-1}\left|\hat{\mathrm{F}}_n - \mathrm{F}_n\right|^2\right]\\&=  \frac{1}{NK^2} \sum_{n=0}^{N-1}\mathbb{E}\left[\left(\sum_{k=0}^{K-1}\left(\mathrm{W}_n{\mathrm{H}}_{k, n} \mathrm{B}_{k, n}-1\right) \mathrm{C}_{k, n} + \mathrm{W}_n \boldsymbol{\Omega}_n\right)^2\right] \\
        &=  \frac{1}{NK^2} \sum_{n=0}^{N-1}\Big( \underbrace{\sum_{k=0}^{K-1}\left|\mathrm{W}_n {\mathrm{H}}_{k, n} \mathrm{B}_{k, n}-1\right|^2}_{\text {Signal misalignment error}} +\underbrace{\mathrm{W}_n^2\sigma_\omega^2}_{\text{Noise-induced error}}\Big).
    \end{aligned}
\end{equation} 
\end{footnotesize}where the first term is the signal misalignment error from the residual channel-gain mismatch, and the second term is the noise-induced error from AP. Removing the constant term in \eqref{obj}, we have $\overline {\mathrm{MSE}} =\sum_{n=1}^{N-1}\left(\sum_{k=0}^{K-1}\left|\mathrm{W}_n {\mathrm{H}}_{k, n} \mathrm{B}_{k, n}-1\right|^2 +\left\|\mathrm{W}_n\right\|^2 \sigma_\omega^2\right).$

\begin{figure*}[!t]
\hrule
\begin{tiny}
\begin{align}
&\mathcal{L} = \sum_{n=1}^{N} \left( \left| \mathrm W_n \mathrm{H}_{k,n} \right| \mathcal{B}_{k,n} - 1 \right)^2 + \frac{1}{2\delta_k} \left[ \left( \max \left\{ 0, \lambda_k + \delta_k \left( \sum_{n=1}^{N} \mathcal{B}_{k,n}^2 - P_\mathrm{t} \right) \right\} \right)^2 - \lambda_k^2 \right] + \frac{1}{2\beta_k} \left[ \left( \max \left\{ 0, \mu_k + \beta_k \left( \left(\sum_{n=1}^{N} \mathcal{B}_{k,n}^2\right)^{-1} - \rho_k \right) \right\} \right)^2 - \mu_k^2 \right]\label{lagrangian1}.\\
&\mathcal{B}_{k,n}^{t+1}= \left( \left| \mathrm W_n \mathrm{H}_{k,n} \right|^2 + \left[ \lambda_k + \delta_k \left( \sum_{n=1}^{N} \mathcal{B}_{k,n}^2 - P_\mathrm{t} \right) \right] 2\delta_k - 2\beta_k\left[ \mu_k + \beta_k \left( \left(\sum_{n=1}^{N} \mathcal{B}_{k,n}^2\right)^{-1} - \rho_k \right) \right] \left( \sum_{n=1}^{N} \mathcal{B}_{k,n}^2 \right)^{-2}  \right) ^{-1}\mathrm W_n \mathrm{H}_{k,n}\label{B_kn}.
\end{align}
\end{tiny}
\end{figure*}
\subsubsection{Radar Sensing Model}
Here, we aim to derive the CRLB for the estimation of Doppler and delay using \eqref{ofdm}. Firstly, we need to obtain the received echoes at each device. 
Three types of channels relevant to device $k$ are considered: (i) the target response channel, characterized by a memory length of $L_\mathrm{trc}$ and impulse response ${g_{k,1}, \dots, g_{k,L_\mathrm{trc}}}$; (ii) the interference channel from a neighboring device $j$, denoted as ${g_{j,1}, \dots, g_{j,L_\mathrm{ic}}}$ with memory $L_\mathrm{ic}$; and (iii) the direct device channel from device $j$ to $k$, represented by ${g_{j,1}^{'}, \dots, g_{j,L_\mathrm{ddc}}^{'}}$ of length $L_\mathrm{ddc}$ \cite{li2023integrated}.
Hence, the target reflection signal received at the $k$-th device can be expressed as
\begin{equation}\label{10}    u_{k,s}=\sum_{l=0}^{L_\mathrm{trc}-1}{g}_{k,l}x_{k,s-\tau_{k,l}}e^{j2\pi T_of_d}e^{j\psi}+\phi_{k,s}+\bar{z}_{k,s}, 
\end{equation} 
where $\tau_{k,l}$ denotes the round-trip delay, $T_o$ denotes OFDM symbol duration, $f_d$ denotes Doppler shift, $\psi$ denotes random phase noise,  $\phi_{k,s}$ denotes the interference signal as \eqref{11}\cite{li2023integrated} and 
$\bar{z}_{k,s} \in\mathbb{C}$ is the AWGN with distribution $\mathcal{CN}(0,\sigma_z^2)$, which is statistically independent of $x_{k,s}$, wherein 
\begin{footnotesize}
 \begin{equation}\label{11}
    \phi_{k,s}=\sum_{i=0,i\neq k}^{K-1}\sum_{l=0}^{L_\mathrm{ic}-1}{g}_{j,l}x_{i,s-\tau_{i,l}}+\sum_{i=0,i\neq k}^{K-1}\sum_{l=0}^{L_\mathrm{ddc}-1}{g}^{'}_{j,l}x_{i,s-\tau_{i,l}}.
\end{equation}   
\end{footnotesize}

 \begin{assumption}\label{assumption}
    \label{assumption1}
    The relative time gaps between any two multipaths are very small in comparison to the actual roundtrip delays, i.e., $\tau_{k,l}=\tau_{k,0},\forall l$ \cite{sen2010adaptive}. Similarly, the attenuation coefficient gaps are also very small \footnote{This assumption can be justified in systems where the path lengths of multipath arrivals differ little (e.g., narrow urban canyon where the range is much greater than the width).}.
\end{assumption}

Based on \textbf{Assumption \ref{assumption1}} and matched filtering, \eqref{10} can be simplified as \textbf{Theorem \ref{theorem1}}. Specifically, the matched filter of length $P$ is defined as $\mathcal{H}_{k,s} = c^*_{k,P-s}$, where $c^*_{k,s}$ is derived by applying the IDFT to the normalized measured data $\mathrm{C}^*_{k,n}$. It can be proved that $c^*_{k, s}$ is independent with unit variance and zero mean, i.e., $\mathbb{E}[c_{k_1,s}c^*_{k_1,s}] = 1, \mathbb{E}[c_{k_1,s}c^*_{k_2,s}] = 0,\forall s\in {\mathcal{N}},k_1,k_2 \in \mathcal{K},k_1 \neq k_2$.

\begin{theorem}\label{theorem1}
After matched filtering and DFT, the Doppler and the delay
can be decoupled and the received signal can be approximated as 
\begin{scriptsize}
\begin{equation} \label{A_{k,n}}
    \mathrm A_{k,n} = \alpha \mathrm B_{k,n} e^{j 2 \pi T_o f_d} e^{-j \frac{2\pi n (\tau_{k}+P)}{N}}e^{j\psi} + \mathrm Z_{k,n}, \forall k \in \mathcal{K}, n \in \mathcal{CN},
\end{equation}    
\end{scriptsize}where $\alpha$ denotes subcarrier attenuation coefficient and $\mathrm Z_{k,n}$ denotes the AWGN with distribution $\mathcal{CN}(0,\sigma_z^2)$.
\end{theorem}
\begin{IEEEproof}
 Please refer to Appendix \ref{theorem1_proof}.
\end{IEEEproof}

\begin{theorem}\label{theorem0}
The CRLBs on the estimation MSEs of distance ${d_{k}}$ and velocity ${v_{k}}$ can be approximated by
\begin{footnotesize}
  \begin{align}
&{\mathop{\rm var}} ( {\hat{d}_k }) \ge\frac{{3 \sigma_z^2 c_o^2}}{8\pi^2\Delta f^2{{\|\mathbf B_k\|}^2\alpha^2 {M}{N}\left( {N^2- 1} \right)}} =\mathrm{CRLB}(\hat{d}_k),\label{31}\\
&{\mathop{\rm var}} ( {\hat { v}_k }) \ge\frac{{3 \sigma_z^2 c_o^2}}{8\pi^2T_o^2f_c^2{{\|\mathbf B_k\|}^2\alpha^2 {M}{N}\left( {M^2 - 1} \right)}}=\mathrm{CRLB}(\hat{v}_k)\label{32}.
\end{align}  
\end{footnotesize}
 
\end{theorem}
\begin{IEEEproof}
 Please refer to Appendix \ref{theorem0_proof}.
\end{IEEEproof}

Given the sensing MSE threshold $\eta_k$ and $\xi_k$, the sensing quality requirement of the $k$-th device is $\mathrm{CRLB}(\hat{d}_k)\le \eta_k$ and $\mathrm{CRLB}(\hat{v}_k)\le \xi_k$.
Let $\eta'_k = \frac{{3 \sigma_z^2 c_o^2}}{8\eta_k\pi^2\Delta f^2{\alpha^2 {M}{N}\left( {N^2- 1} \right)}}$, $\xi'_k = \frac{{3 \sigma_z^2 c_o^2}}{8\xi_k\pi^2T_o^2f_c^2{\alpha^2 {M}{N}\left( {M^2 - 1} \right)}}$, $\rho'_k = \max(\eta'_k, \xi'_k)$ and $\rho_k = \frac{1}{\rho'_k}$, the CRLB constraint can be equivalently derived as $||\mathbf{B}_k||^{-2}\leq \rho_k,\forall k\in\mathcal{K}$ \footnote{This letter focuses on the power control at transmitter and receiver under a given CP length $T_c$, data symbols duration $T$ and subcarrier spacing $\Delta f$. When increasing the subcarrier spacing $\Delta f$ and symbol duration $T_o$, it is advantageous to decrease CRLB.}.

\subsection{Problem Formulation}\label{Formulation}
Our goal is to jointly optimize the transmitting vector $\mathbf{B}_{k}$ and the aggregation vector $\mathbf{W}$ to minimize the MSE under the constraints of maximum power and CRLB. Mathematically, the optimization problem is formulated as

\begin{subequations}\label{problem}
 \begin{align}
(\mathrm{P}1) \min_{\{\mathrm B_{k, n}, \mathrm{W}_n\}_{k \in \mathcal{K}}^{n \in \mathcal{N}}}
    &\quad\overline {\mathrm{MSE}} \label{mse_obj}\\\text{subject to}
&\quad ||\mathbf{B}_k||^2\leq P_\mathrm{t}, \forall k\in\mathcal{K},\label{eq:constraint1}\\
&\quad ||\mathbf{B}_k||^{-2}\leq \rho_k, \forall k\in\mathcal{K}.
\label{eq:constraint2}
\end{align}
 \end{subequations}
The problem is non-convex due to the coupling between ${\mathrm B_{k, n}}$ and $\mathrm{W}_n$ in the objective function. Moreover, the inequality constraint \eqref{eq:constraint2} renders the feasible region non-convex. Therefore, it is challenging to obtain the globally optimal solution for $(\mathrm{P}1)$. Furthermore, both constraints cannot be simultaneously active. To ensure the feasibiliby of the problem, we have  $\rho_k \leq P_{\mathrm{t}}$.
 \begin{figure}[!h]
	\centering
	{\includegraphics[width=0.5\textwidth]{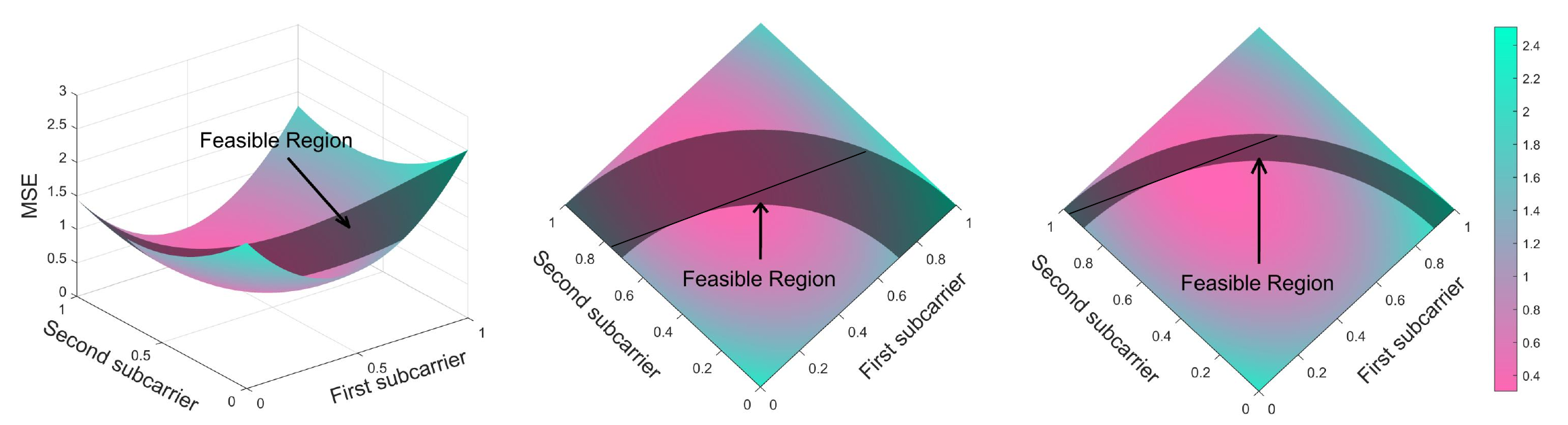}}
	\caption{Visualization of the feasible region in the SCA-based AO phase, taking a two-subcarrier system as an example.}
        \label{feasible}
\end{figure}
\vspace{-0.5cm}
\section{Proposed Solution}
In this section, we propose a two-phase iterative optimization algorithm to solve $(\mathrm{P}1)$. This algorithm consists of two phases: SCA-based AO phase and ADMM-based refinement phase.
 \begin{remark}
    \label{remark0}
    The choice of the two-phase iterative algorithm is driven by the mathematical structure of $(\mathrm{P}1)$. Specifically, this problem involves non-convex coupling between the transmit and aggregation variables, as well as a non-convex feasible region due to the CRLB constraint. Given this structure, decoupling the variables and solving the subproblems iteratively is a natural approach in non-convex optimization.
\end{remark}
\subsection{SCA-Based Alternating Optimization Phase}
Alternating optimization can be used to decompose $(\mathrm{P}1)$ into two more tractable subproblems \footnote{Alternating optimization aims to realize a trade-off between signal misalignment error and noise-induced error in Eq. \eqref{obj}. It is similar to the trade-off between the zero-forcing precoder and matched-filter receiver.}. Initially, we align the phase as $\mathrm B_{k, n} = \mathcal B_{k, n} \times \frac{{\mathrm{H}}_{k, n} \mathrm{W}_n}{\left |\mathrm{W}_n {\mathrm{H}}_{k, n}\right |}$ to minimize the MSE \cite{chen2023over}, where $\mathcal B_{k, n}$ denotes the amplitude. The first subproblem focuses on optimizing $\mathrm{W}_n$ given $\mathrm B_{k, n}$, which can further be partitioned into the following $N$ subproblems
\begin{equation}
     \begin{aligned}
    (\mathrm{P}1.1) \min_{\{\mathrm{W}_n\}}
        &\quad \sum_{k =0}^{K-1}\left|\mathrm{W}_n {\mathrm{H}}_{k, n} \mathrm B_{k, n}-1\right|^2 + \mathrm{W}_n^2 \sigma_\omega^2.  
    \end{aligned}
\end{equation}
It is an unconstrained problem. The optimal $\mathrm{W}_n^{\star}$ can be obtained according to first-order condition as
\begin{equation}\label{W_n}
\mathrm{W}_n^{\star} = \left(\sum_{k =0}^{K-1}\left|\mathrm B_{k, n}\right|^2 \left|\mathrm{H}_{k, n}\right|^2 +\sigma_\omega^2\right)^{-1} \sum_{k =0}^{K-1} {\mathrm{H}}_{k, n} \mathrm B_{k, n}.
\end{equation}

The second subproblem aims to optimize $\mathcal B_{k, n}$ given $\mathrm{W}_n$, which can be decomposed into the following $K$ subproblems.

\begin{equation}
     \begin{aligned}
    (\mathrm{P}1.2) \min_{\{\mathrm B_{k, n}\}}
        &\quad \sum_{n =0}^{N}\left(\left|\mathrm{W}_n {\mathrm{H}}_{k, n}\right| \mathcal B_{k, n}-1\right)^2 + \mathrm{W}_n^2 \sigma_\omega^2\label{}  \\\text{subject to}
&\quad \eqref{eq:constraint1},\eqref{eq:constraint2}. 
    \end{aligned}.
\end{equation}

When sensing constraints of $(\mathrm{P}1.2)$ are removed, this problem becomes a quadratically constrained quadratic programming (QCQP) problem, which is a convex problem. Let $\mu_k$ denote the dual variable associated with the transmit power constraint, we have the following theorem.
\begin{theorem}\label{theorem2}
    When $P_{\mathrm{t}} \to \infty$, we have $\mathrm{W}_n^{\star} \to 0$ and thus the asymptotic $\overline {\mathrm{MSE}} \to 0$ (lower bound).
\end{theorem}
\begin{IEEEproof}
 Please refer to Appendix \ref{theorem2_proof}.
\end{IEEEproof}

When considering the sensing constraint, the problem becomes non-convex. The non-convexity stems from the non-convex feasible region. To tackle this issue, we use the SCA method based on the first-order Taylor expansion as 
\begin{equation}
\left(||\mathbf B_k^i||^2 + 2\mathbf B_k^i\left(\mathbf B_k-\mathbf B_k^i\right)\right)^{-1} \leq \rho_k,\forall k\in\mathcal{K},
\end{equation}
where $\mathbf B_k^i$ represents a feasible reference point on the boundary of CRLB constraint. Moreover, $||\mathbf B_k^i||^2 + 2\mathbf B_k^i(\mathbf B_k-2\mathbf B_k^i)$ acts as an upper bound for $||{\mathbf B}_k||^2$. Then, this subproblem becomes a QCQP, which can be solved by the existing toolbox, such as CVX. 

Finally, $(\mathrm{P}1)$ is relaxed into two separate convex subproblems, whose feasible region is a subset of that of $(\mathrm{P}1)$. 

 \begin{remark}
    \label{remark1}
    Although SCA-based AO phase can obtain a suboptimal solution, it sacrifices part of the feasible region due to the heuristic initial reference point. As is shown in Fig. \ref{feasible}, the proportion of sacrifice increases, when $P_\mathrm{t}$ and $\rho_{k}$ get closer. When $P_{\mathrm{t}} \to \infty$ or CRLB $\to 0$ (e.g., $N \to \infty$), the feasible region sacrifice can be neglected.
\end{remark}

\begin{algorithm}[h]
\caption{Two-Phase Iterative Optimization Algorithm.}
\label{alg1}
\begin{algorithmic}[1]
\begin{footnotesize}
    \STATE{\textbf{Initialize} $\mathcal{B}_{k, n}^0, \mathbf{W}^0, \lambda^0, \mu^0, \delta^0, \beta^0, \epsilon_{\text{mse}}, \epsilon_{\text{pc}}, \epsilon_{\text{sc}}$.}
\STATE{\textbf{Output} $\mathcal{B}_{k, n}^{\star}, \mathbf{W}^{\star}$.}
\STATE{\textbf{**SCA-Based Alternating Optimization Phase**}}
\REPEAT
    \STATE{Update $\mathbf{W}$ using \eqref{W_n};}
    \STATE{Update $\mathbf{B}_k$ with fixed $\mathbf{W}$}
\UNTIL{the reduction in MSE is less than $\epsilon_{\text{MSE}}$}
\STATE{\textbf{**ADMM-Based Refinement Phase**}}
\FOR{$k = 1$ to $K$}
    \REPEAT
        \STATE{Update $\mathbf{\mathcal{B}}_{k,n}^{t+1}$ using \eqref{B_kn}.}
        \STATE{Update $\lambda_k^{t+1}, \mu_k^{t+1}, \delta_k^{t+1}, \beta_k^{t+1}$ using \eqref{lambda_k}.}
        \STATE{Update $\delta_{t+1}, \beta_{t+1}$ using \eqref{delta_k}.}
        \STATE{$t = t + 1$.}
    \UNTIL{\small the reduction in MSE is less than $\epsilon_{\text{MSE}}$ and $\Gamma_{pv} > \epsilon_{\text{pc}}$ and $\Gamma_{sv} > \epsilon_{\text{sc}}$}
\ENDFOR
\end{footnotesize}

\end{algorithmic}
\end{algorithm}
\vspace{-0.5cm}
\subsection{ADMM-Based Refinement Phase}
To search a broader feasible region and further improve the MSE, we adopt the ADMM method to refine $\mathcal B_{k, n}$ using the initial solution obtained from SCA-based AO phase \footnote{ADMM is particularly suitable for problems with complicated constraints, as it enables distributed optimization \cite{conejo2006decomposition}. Here, ADMM-based refinement focuses on the sensing-computation trade-off by refining $\mathcal B_{k, n}$. }. We use additional variables to transform inequality constraints into equality constraints \cite[pp. 205-207]{conejo2006decomposition}. Then, the augmented Lagrangian function can be obtained as \eqref{lagrangian1}, where $\lambda_k$ and $\mu_k$ denote the Lagrangian dual variables associated with the constraints \eqref{eq:constraint1} and \eqref{eq:constraint2}, respectively, and $\delta_k$ and $\beta_k$ denote the penalty parameters. At the $(t +1)$-th iteration, the ADMM algorithm consists of the primal updates, the dual updates and the penalty update steps. In such a way, we have

\begin{footnotesize}
 \begin{equation}\label{primal}
  \mathcal{B}_{k,n}^{t+1} = \operatorname*{argmin}_{\mathcal{B}_{k,n}^t} \mathcal{L}(\mathcal{B}_{k,n}^t,\lambda_k^t, \mu_k^t, \delta_k^t, \beta_k^t),
\end{equation}   
\end{footnotesize}which can be solved by checking the first-order optimality conditions and its solution is given by \eqref{B_kn}, with
\begin{footnotesize}
    \begin{equation}\label{lambda_k} 
    \begin{aligned}
        \lambda_k^{t+1} &= \max\left(0, \lambda_k^t + \delta_k^t \left( \left\lVert \mathcal{B}_k \right\rVert_2^2 - P_k \right) \right),\\
        \mu_k^{t+1} &= \max\left(0, \mu_k^t + \beta_k^t \left( \left\lVert \mathcal{B}_k \right\rVert_2^{-2} - \rho_k \right) \right),
    \end{aligned}
\end{equation}
\end{footnotesize}Since the dual variables are non-negative, the max function is employed. In addition, it is impossible for both $\lambda_k$ and $\mu_k$ to be zero simultaneously. This is because, as stated in Section \ref{Formulation}, the two constraints cannot be active simultaneously. Finally we update the penalty as
\begin{small}
\begin{equation}\label{delta_k} 
    \delta_k^{t+1} = 
    \begin{cases}
    \mathcal{G} \delta_k^{t}, & \Gamma_{pv} > \epsilon_{pc} \\
    \mathcal{D}\delta_k^{t}, & \Gamma_{pv} \leq \epsilon_{pc} \\
    \delta_k^{t}, & \text{otherwise}
    \end{cases}, 
    \beta_k^{t+1} = 
    \begin{cases}
    \mathcal{G} \beta_k^{t}, & \Gamma_{sv} > \epsilon_{sc} \\
    \mathcal{D}\beta_k^{t}, & \Gamma_{sv} \leq \epsilon_{sc} \\
    \beta_k^{t}, & \text{otherwise}
    \end{cases},
\end{equation}
\end{small}where $\mathcal{G}$ and $\mathcal{D}$ denote the growth and decay coefficients, $\Gamma_{pv}$ and $\Gamma_{sv}$ denote the violations of the power constraint and the sensing constraint, and $\epsilon_{pc}$ and $\epsilon_{sc}$ denote the maximum thresholds of the constraint violations.

 \begin{figure*}[!t]
	\centering
        \begin{minipage}{0.23\textwidth}
		{\includegraphics[width=\textwidth]{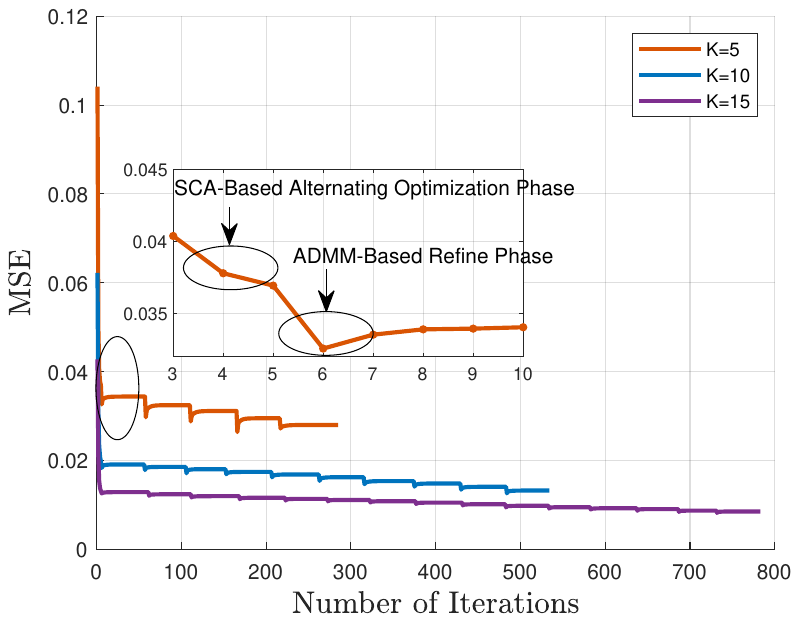}} 
            \caption*{\footnotesize (a)}\label{Fig_LOSS}
	\end{minipage}
	\begin{minipage}{0.23\textwidth}
		{\includegraphics[width=\textwidth]{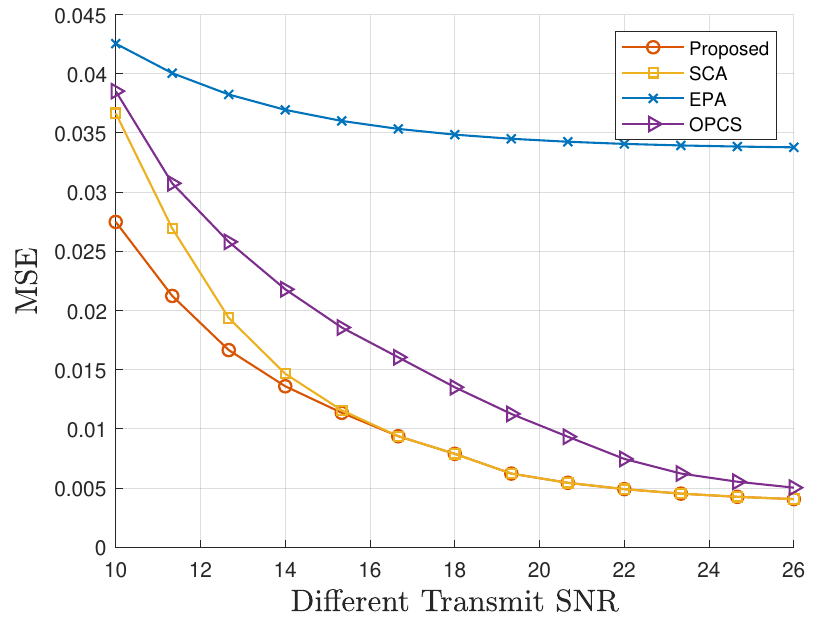}}
		\caption*{\footnotesize (b)}\label{Fig_P}
	\end{minipage}
        \begin{minipage}{0.23\textwidth}
		{\includegraphics[width=\textwidth]{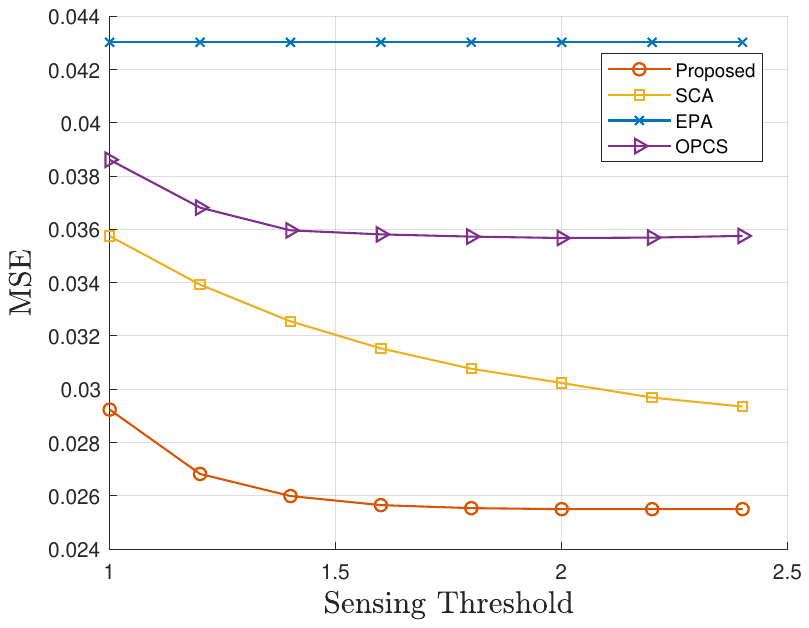}}
		\caption*{\footnotesize (c)}\label{Fig_L}
	\end{minipage}
        \begin{minipage}{0.23\textwidth}
		{\includegraphics[width=\textwidth]{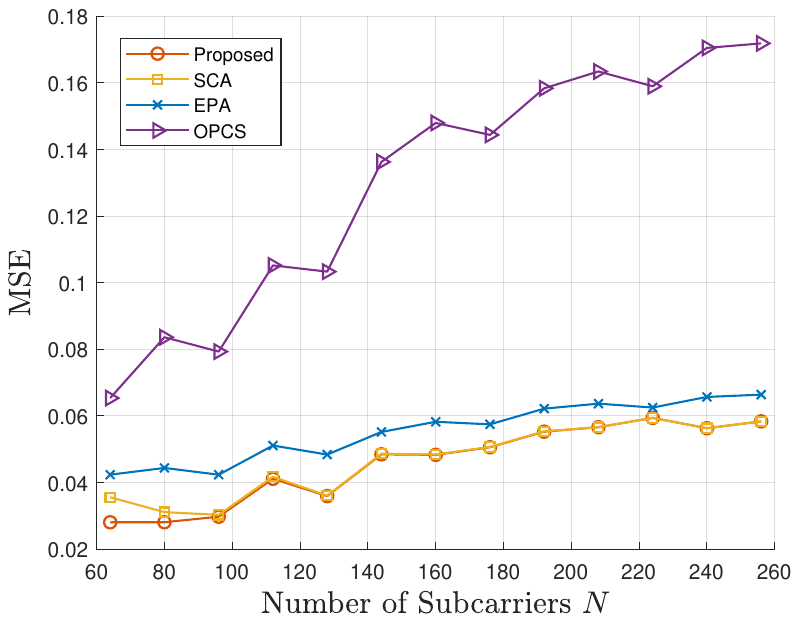}}
		\caption*{\footnotesize (d)}\label{Fig_N}
	\end{minipage}
\caption{(a) Convergence performance of \textbf{Algorithm \ref{alg1}}. (b) MSE versus the transmit power. (c) MSE versus the sensing thresholds.  (d) MSE versus the number of subcarriers.}\label{Fig}
\end{figure*}

The details of the proposed method are summarized in \textbf{Algorithm \ref{alg1}}. In each iteration of AO phase, the MSE decreases or remains unchanged since each subproblem is convex in terms of ${\{\mathrm B_{k, n}\}}$ and ${\{\mathrm{W}_n\}}$. Meanwhile, the ADMM phase can converge locally to the solution at a linear rate \cite{conejo2006decomposition}. The complexity analysis of \textbf{Algorithm \ref{alg1}} is given as follows. The complexity for SCA-based AO phase is dominated by QCQP with a worst-case complexity $\mathcal{O}(K^{4.5}N^{4.5} \mathrm{log}(1/\epsilon))$, where a prefixed solution accuracy is $\epsilon$ \cite{huang2014randomized}. As for the ADMM-based refinement phase, the calculation of $\mathcal{B}_{k,n}$ has a complexity of $\mathcal{O}(KN)$.

\section{Numerical Results}
In this section, the numerical results are given to evaluate the performance of our proposed method. In the simulation, we set $P_\mathrm{t} = 10$ dBm, $\sigma_\omega^2=-20$ dBm, $\eta_k$ = 1, $\xi_k$ = 1, $K = 5$, $M = 50$, $N = 64$, $T_o = 8$ $\mathrm{\mu s}$, and $\Delta f = 156.25$ $\mathrm{kHz}$.



For comparison, we consider the following benchmarks: (i) equal power allocation (EPA) $\mathcal B_{k, n}= \frac{\sqrt{P_\mathrm{t}}}{N}$; (ii) optimal power allocation for each subcarrier (OPAS) $\mathcal B_{k, n}=\mathrm{max}(\sqrt{\frac{\rho'_k}{N}}, \mathrm{min}(\sqrt{\frac{P_\mathrm{t}}{N}},\frac{1}{\left |\mathrm{W}_n {\mathrm{H}}_{k, n}\right |}))$ \cite{liu2020over}; (iii) SCA-based power allocation. \footnote{The proposed ISCC system is different from existing works. Hence, existing state-of-the-art methods in the ISAC and AirComp literature cannot be directly applied to solve the same optimization problem. In this regard, meaningful comparisons with these methods are currently not feasible.}

Fig. \ref{Fig}(a) shows the convergence performance of the proposed \textbf{Algorithm \ref{alg1}}. We can observe that the convergence of the algorithm is divided into two phases. There are $K$ distinct falling edges in the ADMM-Based refinement phase, which corresponds to a device-by-device optimization given 
$\mathrm{W}_n^{\star}$. Additionally, as $K$ increases, the lower bound of the convergent MSE becomes smaller. It is equivalent to the convergence in the scenario of federated learning (FL) utilizing deep neural network (DNN) training \cite{wen2024survey}.

Fig. \ref{Fig}(b) shows the MSE versus transmit power $P_\mathrm{t}$. It is observed that the proposed method outperforms the baselines. In the low SNR region, the proposed method enjoys a refinement gain compared to SCA. As the transmit power increases, the proposed method can approach the MSE lower bound more effectively. Meanwhile, the MSE gap between OPAS, SCA and the proposed method becomes less. This verifies \textbf{Remark \ref{remark1}}. 

Fig. \ref{Fig}(c) demonstrates the performance trade-off between computation and sensing. The results are obtained by setting different sensing thresholds. It is evident that the EQA remains unaffected by the sensing thresholds because it simply divides the maximum power equally. However, the MSE decreases as the sensing threshold rises in the other methods. The proposed method achieves a better sensing-computation trade-off compared to the baselines. 

Fig. \ref{Fig}(d) shows the MSE versus the number of subcarriers $N$. From this figure, we have three observations. Firstly, regardless of the value of $N$, the proposed method outperforms the baselines. Secondly, as $N$ increases, the performance of the OPCS deteriorates. This is because noise-induced error dominates MSE when N increases and 
it is difficult to design one common $\mathbf{W}$ for all subcarriers. Finally, the MSE of the proposed method initially shows fluctuations along with the increase of $N$. The reason is that $N$ has a dual impact on the MSE: (i) it affects the size of the feasible region, thereby influencing the quality of the initial solution in SCA-Based AO phase; (ii) designing a common $\mathbf{W}$ becomes more challenging as $N$ increases. This also verifies \textbf{Remark \ref{remark1}}.

\section{Conclusion}
 In this letter, the issue of jointly designing transmitting vector and aggregation vector is addressed for the OFDM-based ISCC system. A two-phase iterative optimization algorithm is proposed to optimize transceiver vectors and minimize the computational MSE under CRLB and power constraints. In the first phase, an SCA-based AO algorithm is adopted to decouple transceiver vectors and relax the feasible region. Based on the obtained solutions, we further improve MSE by refining transmitting vector under the full feasible region. Simulations and analysis have shown that the proposed algorithm can achieve a better sensing-computation trade-off.
\appendix
\subsection{Proof of Theorem \ref{theorem1}}\label{theorem1_proof}
Firstly, matched filtering is employed under $M$ observed OFDM symbols as
\begin{scriptsize}
\begin{equation} \label{25}
    \begin{aligned}
    &\mathbb{E}[u_{k,s}]=\mathbb{E}\Big[\Big(\sum_{l=0}^{{L}_\mathrm{trc}-1}{g}_{k,l}x_{k,s-\tau_{k,l}}e^{j2\pi T_of_d}e^{j\psi}+\phi_{k,s}+\bar{z}_{k,s}\Big)*\mathcal{H}_{k,s}\Big]
        \\&\stackrel{(a)}{\approx}\frac{1}{M}\sum_{m=0}^{M-1}\Big(\sum_{l=0}^{{L}_\mathrm{trc}-1}{g}_{k,l}x_{k,s-\tau_{k,0}}e^{j2\pi T_of_d}e^{j\psi}+\phi_{k,s}+\bar{z}_{k,s}\Big)*\mathcal{H}_{k,s}\\
     & \stackrel{(b)}{\approx} \frac{1}{M}\sum_{m=0}^{M-1}\Big[\sum_{p=0}^{P-1}\Big(\sum_{l=0}^{{L}_\mathrm{trc}-1}{g}_{k,l}x_{k,s-p-\tau_{k,0}}e^{j2\pi T_of_d}e^{j\psi}\Big)c^*_{k,P-p}\Big] + {z}_{k,s}\\
     & = u_{k,s}^{'},
    \end{aligned}
\end{equation}
\end{scriptsize}where ${z}_{k,s}=\frac{1}{M}\sum_{m=0}^{M-1}\sum_{p=0}^{P-1}\bar{z}_{k,j}c^*_{k,P-p}$ denotes the AWGN with distribution $\mathcal{CN}(0,\sigma_z^2)$; $(a)$ is founded on the law of large-number when $M$ is large and \textbf{Assumption \ref{assumption1}}; $(b)$ follows that $c^*_{k,s}$ is independent with unit variance and zero mean.

Then, by applying the DFT to \eqref{25}, we have
\begin{footnotesize}
\begin{equation}\label{after_mf}
    \begin{aligned}
        \mathrm A_{k,n} &= \frac{1}{\sqrt N} \sum_{s=0}^{N-1} u_{k,s}^{'}e^{-j \frac{2\pi n s}{N}} \\
        &\stackrel{(a)}{\approx} \mathrm{G}_{k,n} \mathrm B_{k,n} e^{j 2 \pi T_o f_d} e^{-j \frac{2\pi n (\tau_{k,0}+P)}{N}}e^{j\psi} + \mathrm Z_{k,n}.
    \end{aligned}
\end{equation}\end{footnotesize}where $\mathrm{Z}_{k,n}=\frac{1}{\sqrt{N}}\sum_{s=0}^{N-1}\sum_{l=0}^{{L}_\mathrm{trc}-1}{z}_{k,s}e^{-j\frac{2\pi}{N} sn}$ denotes the AWGN with distribution $\mathcal{CN}(0,\sigma_z^2)$ and $\mathrm{G}_{k,n}=\frac{1}{\sqrt{N}}\sum_{s=0}^{N-1}\sum_{l=0}^{{L}_\mathrm{trc}-1}{g}_{k,l}e^{-j\frac{2\pi}{N} sn}$ denotes the subcarrier attenuation coefficient. Based on \textbf{Assumption \ref{assumption1}}, we have $\mathrm{G}_{k,n}=\mathrm{G}_{k,0}, \forall n\in\mathcal{N}$. The $\mathrm{G}_{k,0}$ is equivalent to a constant coefficient $\alpha =\sqrt{\frac{c_0^2\sigma_{\mathrm{RCS}}}{(4\pi)^3d_k^4f_{c}^2}}$ \cite{braun2014ofdm}, where $\sigma_{\mathrm{RCS}}$ denotes the radar cross section, $d_k$ is the distance between the $k$-th device and the target, $c_0$ denotes the speed of light and $f_{c}$ denotes the carrier frequency. Additionally, $(a)$ follows $c^*_{k, s}$ is with unit variance and the assumption that $f_d$ is far less than subcarrier spacing $\Delta f$ \cite{gaudio2019performance}. \eqref{after_mf} shows that doppler shift and the delay are decoupled. 

 \vspace{-0.5cm}
\subsection{Proof of Theorem \ref{theorem0}}\label{theorem0_proof}
For the convenience of analysis, we denote $\bar \tau = 2\pi\Delta f \tau_{k,0}$ and $\bar v = 2 \pi T_o f_d$.
Let $\boldsymbol{\theta}= {\left[ { \bar \tau ,\bar v, \alpha, \psi } \right]^T}$ and $\mathcal{A}_{k,n}(\boldsymbol{\theta})$ denote the log-likelihood function of $\mathrm{A}_{k,n}$, the second-order Fisher information matrix is 
\begin{footnotesize}
   \begin{align}
{{\mathcal{J}}_{ij}} = \frac{1}{{ \sigma_z^2}}\sum\limits_{m = 0}^{{M} - 1} {\sum\limits_{n = 0}^{{N} - 1} {{\left[ {\frac{{\partial \mathcal A_{k,n}^{{\mathop{\rm Re}\nolimits} }}}{{\partial {\theta_i}}}\frac{{\partial \mathcal A_{k,n}^{{\mathop{\rm Re}\nolimits} }}}{{\partial {\theta_j}}}{\rm{ + }}\frac{{\partial \mathcal A_{k,n}^{{\mathop{\rm Im}\nolimits} }}}{{\partial {\theta_i}}}\frac{{\partial \mathcal A_{k,n}^{{\mathop{\rm Im}\nolimits} }}}{{\partial {\theta_j}}}} \right]} } },\label{eqappendix2}
 \end{align}  
\end{footnotesize}where $\theta_i$ is the $i$-th entry of vector $\boldsymbol{\theta}$, upper subscript $\mathop{\rm Re}$ and $\mathop{\rm Im}$ denote the real and imaginary parts of $\mathcal A_{k,n}$. 

Then, we can derive CRLB with the inverse of the above Fisher information matrix as
\begin{footnotesize}
 \begin{align}
{\mathop{\rm var}} \left( {{{\hat \theta}_i}} \right) \ge {{\mathcal{J}}^{ - 1}}\left( {i,i} \right)\label{eqcrb}.
 \end{align}\end{footnotesize}Given the complexity of the inverse processing, it is difficult to derive the closed-form CRLB from \eqref{eqcrb}. Here, we use averaged CRLB provided by \cite[chapter 3.3]{braun2014ofdm}, which is based on the fact that each OFDM symbol can be used for one estimation and the average of all estimations is unbiased. Hence, we have 
\begin{footnotesize}
 \begin{align}
{\mathop{ \rm var}} \left( {\hat {\theta}_i } \right) \ge \frac{1}{{{M}{N}}}{\mathcal{J}}_{{\theta}_i}^{ - 1}\left( {i,i} \right),\label{eqappendix4}
 \end{align}\end{footnotesize}where
${\mathcal{J}}_{\theta_1} = \frac{1}{{ \sigma_z^2}}\sum\nolimits_{n = 0}^{{N} - 1} {\left[ {\frac{{\partial \mathcal A_{k,n}^{{\mathop{\rm Re}\nolimits} }}}{{\partial {\theta_i}}}\frac{{\partial \mathcal A_{k,n}^{{\mathop{\rm Re}\nolimits} }}}{{\partial {\theta_j}}}{\rm{ + }}\frac{{\partial \mathcal A_{k,n}^{{\mathop{\rm Im}\nolimits} }}}{{\partial {\theta_i}}}\frac{{\partial\mathcal A_{k,n}^{{\mathop{\rm Im}\nolimits} }}}{{\partial {\theta_j}}}} \right]}, i, j \in \left\{ {1,3,4} \right\}$ and ${\mathcal{J}}_{\theta_2} = \frac{1}{{ \sigma_z^2}}\sum\nolimits_{n = 0}^{{M} - 1} {\left[ {\frac{{\partial \mathcal A_{k,n}^{{\mathop{\rm Re}\nolimits} }}}{{\partial {\theta_i}}}\frac{{\partial \mathcal A_{k,n}^{{\mathop{\rm Re}\nolimits} }}}{{\partial {\theta_j}}}{\rm{ + }}\frac{{\partial \mathcal A_{k,n}^{{\mathop{\rm Im}\nolimits} }}}{{\partial {\theta_i}}}\frac{{\partial\mathcal A_{k,n}^{{\mathop{\rm Im}\nolimits} }}}{{\partial {\theta_j}}}} \right]},$\\$i ,j \in \left\{ {2,3,4} \right\}$. By substituting \eqref{A_{k,n}} into \eqref{eqappendix4}, we have
\begin{footnotesize}
 \begin{align}
{\mathop{\rm var}} \left( {\hat {\bar \tau} }\right) \ge\frac{{6 \sigma_z^2}}{{{\|\mathbf B_k\|}^2\alpha^2 {M}{N}\left( {N^2- 1} \right)}},\label{eqaooendix66}\\
{\mathop{\rm var}} \left( {\hat {\bar v} }\right) \ge\frac{{6 \sigma_z^2}}{{{\|\mathbf B_k\|}^2\alpha^2 {M}{N}\left( {M^2 - 1} \right)}}\label{eqaooendix67}.
\end{align}\end{footnotesize}

Since $d_k=\frac12c_0\tau_{k,0}$ and $\tau_{k,0} = \frac{1}{2\pi\Delta f}\bar \tau$, we have ${d}_k=\frac{c_0}{4\pi\Delta f}\bar \tau$ and then \eqref{31}. 
Since $f_d=\frac{2 v_kf_c}{c_0}$ and $\bar v = 4 \pi T_o \frac{ vf_c}{c_0}$, we have \eqref{32}.

\subsection{Proof of Theorem \ref{theorem2}}\label{theorem2_proof}
 The Lagrangian of the second subproblem is 
 \begin{footnotesize}
  \begin{equation}
     \mathcal{L}_1(\mathcal{B}_{k,n}, \mu_k) = \sum_{n=1}^{N} \left( \left| \mathrm W_n \mathrm{H}_{k,n} \right| \mathcal{B}_{k,n} - 1 \right)^2 +  \mu_k \left( \sum_{n=1}^{N} \mathcal{B}_{k,n}^2 - P_\mathrm{t} \right).  
 \end{equation}    
 \end{footnotesize}By leveraging the stationarity of Karush-Kuhn-Tucker (KKT) conditions, the optimal solution is $\mathcal B_{k,n}^\star=\frac{|\mathrm{W}_n^{\star}\mathrm{H}_{k, n}|}{|\mathrm{W}_n^{\star}\mathrm{H}_{k, n}|^2+\mu_k^\star}$, where $\mu_k^\star$ is the optimal dual variable. By leveraging the complementary slackness condition $\mu_k^\star \left( \sum_{n=1}^{N} \mathcal{B}_{k,n}^2 - P_\mathrm{t} \right)=0$, we have $\mu_k^\star \to 0$ and $\mathrm{W}_n^{\star} \to 0$ when $P_{\mathrm{t}} \to \infty$. By substituting them into \eqref{mse_obj}, we have the asymptotic $\overline {\mathrm{MSE}} \to 0$.
\bibliographystyle{IEEEtran}
\bibliography{OFDM_ISCC_wcl}

\begin{thebibliography}{10}
\providecommand{\url}[1]{#1}
\csname url@samestyle\endcsname
\providecommand{\newblock}{\relax}
\providecommand{\bibinfo}[2]{#2}
\providecommand{\BIBentrySTDinterwordspacing}{\spaceskip=0pt\relax}
\providecommand{\BIBentryALTinterwordstretchfactor}{4}
\providecommand{\BIBentryALTinterwordspacing}{\spaceskip=\fontdimen2\font plus
\BIBentryALTinterwordstretchfactor\fontdimen3\font minus \fontdimen4\font\relax}
\providecommand{\BIBforeignlanguage}[2]{{%
\expandafter\ifx\csname l@#1\endcsname\relax
\typeout{** WARNING: IEEEtran.bst: No hyphenation pattern has been}%
\typeout{** loaded for the language `#1'. Using the pattern for}%
\typeout{** the default language instead.}%
\else
\language=\csname l@#1\endcsname
\fi
#2}}
\providecommand{\BIBdecl}{\relax}
\BIBdecl

\bibitem{wen2024survey}
D.~Wen, Y.~Zhou, X.~Li, Y.~Shi, K.~Huang, and K.~B. Letaief, ``A survey on integrated sensing, communication, and computation,'' \emph{IEEE Comm. Surv. Tutor.}, early access, Dec. 23, 2024, doi: 10.1109/COMST.2024.3521498.

\bibitem{chen2023over}
Y.~Chen, H.~Xing, J.~Xu, L.~Xu, and S.~Cui, ``Over-the-air computation in {OFDM} systems with imperfect channel state information,'' \emph{IEEE Trans. Commun.}, vol.~72, no.~5, pp. 2929--2944, 2023.

\bibitem{liu2021cramer}
F.~Liu, Y.-F. Liu, A.~Li, C.~Masouros, and Y.~C. Eldar, ``Cram{\'e}r-rao bound optimization for joint radar-communication beamforming,'' \emph{IEEE Trans. Signal Process.}, vol.~70, pp. 240--253, 2021.

\bibitem{li2023integrated}
X.~Li, F.~Liu, Z.~Zhou, G.~Zhu, S.~Wang, K.~Huang, and Y.~Gong, ``Integrated sensing, communication, and computation over-the-air: {MIMO} beamforming design,'' \emph{IEEE Trans. Wireless Commun.}, vol.~22, no.~8, pp. 5383--5398, 2023.

\bibitem{liu2024ofdm}
F.~Liu, Y.~Zhang, Y.~Xiong, S.~Li, W.~Yuan, F.~Gao, S.~Jin, and G.~Caire, ``{OFDM} achieves the lowest ranging sidelobe under random {ISAC} signaling,'' \emph{arXiv preprint arXiv:2407.06691}, 2024.

\bibitem{shao2022semantic}
Y.~Shao and D.~Gunduz, ``Semantic communications with discrete-time analog transmission: A {PAPR} perspective,'' \emph{IEEE Wireless Commun. Lett.}, vol.~12, no.~3, pp. 510--514, 2022.

\bibitem{sen2010adaptive}
S.~Sen and A.~Nehorai, ``Adaptive {OFDM} radar for target detection in multipath scenarios,'' \emph{IEEE Trans. Signal Process.}, vol.~59, no.~1, pp. 78--90, 2010.

\bibitem{conejo2006decomposition}
A.~J. Conejo, E.~Castillo, R.~Minguez, and R.~Garcia-Bertrand, \emph{Decomposition techniques in mathematical programming: {E}ngineering and science applications}.\hskip 1em plus 0.5em minus 0.4em\relax Springer Science \& Business Media, 2006.

\bibitem{huang2014randomized}
Y.~Huang and D.~P. Palomar, ``Randomized algorithms for optimal solutions of double-sided {QCQP} with applications in signal processing,'' \emph{IEEE Trans. Signal Process.}, vol.~62, no.~5, pp. 1093--1108, 2014.

\bibitem{liu2020over}
W.~Liu, X.~Zang, Y.~Li, and B.~Vucetic, ``Over-the-air computation systems: {Optimization}, analysis and scaling laws,'' \emph{IEEE Trans. Wireless Commun.}, vol.~19, no.~8, pp. 5488--5502, 2020.

\bibitem{braun2014ofdm}
K.~M. Braun, ``{OFDM} radar algorithms in mobile communication networks,'' Ph.D. dissertation, Karlsruhe, Karlsruher Institut f{\"u}r Technologie (KIT), Karlsruhe, Germany, 2014.

\bibitem{gaudio2019performance}
L.~Gaudio, M.~Kobayashi, B.~Bissinger, and G.~Caire, ``Performance analysis of joint radar and communication using {OFDM} and {OTFS},'' in \emph{Proc. IEEE Int. Conf. Commun. Workshops (ICC Workshops)}, 2019, pp. 1--6.

\end{thebibliography}

\end{document}